\documentclass{article}

\begin{document}

\title{Stability of capillary surfaces supported by a helical wire}
\author{Jorge A. Bernate$^*$ and David B. Thiessen \\
\\\vspace{6pt} Washington State University, \\ Pullman, WA 99164-2710, USA \\
$^*$ \small{Current institution: Stanford University}}
\maketitle

In this fluid dynamics video, we report on experiments of large aspect ratio capillary surfaces supported by a helical wire. These open channels are potentially useful for fluid management at low Bond numbers. The experiments were performed in a Plateau tank, injecting 2-fluorotoluene into distilled water, density matched at 26.5 $^{o}$C.  Extension springs with a nominal diameter of 0.24 inches and with a wire diameter of 0.018 inches were used. At small extensions, liquid can be injected into the spring from one side. The injected free-ended liquid columns are stable. Volume-constrained stability can be studied after injecting from both sides until coalescence. Stable states exist between lower and upper volume stability limits. The low volume instability is manifested by the growth of a neck that eventually breaks. The breakage sends a wave front; the interface ahead of the front is static while the interface behind the front undergoes oscillations. These oscillations eventually dampen as the free-ends retract until equilibrium states are reached. The upper volume instability causes the interface to blow out creating a large drop that remains connected to the liquid column supported by the spring. After a critical pitch, injection is no longer possible from one end. However, a liquid column established by injection can be stretched beyond the critical pitch for injection.  Beyond this pitch, after breakup and blowout the interface undergoes a pearling instability. The resulting free ends retract and grow into drops from which the liquid column eventually breaks. This pearling instability propagates throughout the entire length of the structure largely emptying it. Small tracer drops, outside and in the vicinity of the interface, traced helical trajectories as liquid was  injected into, and withdrawn from, the spring. We also demonstrated vertical wicking into a small spring in air.

\end{document}